\documentclass[10pt,aps,prl,reprint,twocolumn,superscriptaddress,floatfix,showpacs,longbibliography]{revtex4-1}
\usepackage[utf8]{inputenc}
\usepackage[T1]{fontenc}
\usepackage{graphicx}
\usepackage{tabularx}
\usepackage[usenames,dvipsnames,table]{xcolor}
\usepackage[version=3]{mhchem}
\usepackage{braket}
\usepackage{upgreek}
\usepackage{hyperref}
\usepackage{amsmath,amssymb}
\usepackage[normalem]{ulem}
\usepackage{bbm}
\usepackage{soul}

\definecolor{mark}{rgb}{0.85, 0.9, 1}

\hypersetup{hidelinks}
\sethlcolor{mark}

\begin{document}

\title{Achieving the volume-law entropy regime with random-sign Dicke states}



\author{Oleg M. Sotnikov}
\affiliation{Theoretical Physics and Applied Mathematics Department, Ural Federal University, Ekaterinburg 620002, Russia}
\affiliation{Russian Quantum Center, Skolkovo, Moscow 121205, Russia}
\author{Ilia A. Iakovlev}
\affiliation{Theoretical Physics and Applied Mathematics Department, Ural Federal University, Ekaterinburg 620002, Russia}
\affiliation{Russian Quantum Center, Skolkovo, Moscow 121205, Russia}
\author{Evgeniy O. Kiktenko}
\affiliation{Russian Quantum Center, Skolkovo, Moscow 121205, Russia}
\affiliation{National University of Science and Technology ``MISIS'', Moscow 119049, Russia}
\author{Aleksey K. Fedorov}
\affiliation{Russian Quantum Center, Skolkovo, Moscow 121205, Russia}
\affiliation{National University of Science and Technology ``MISIS'', Moscow 119049, Russia}
\author{Vladimir V. Mazurenko}
\affiliation{Theoretical Physics and Applied Mathematics Department, Ural Federal University, Ekaterinburg 620002, Russia}
\affiliation{Russian Quantum Center, Skolkovo, Moscow 121205, Russia}

\date{\today}

\begin{abstract}
Manipulating entanglement, which reflects non-local correlations in a quantum system and defines the complexity of describing its wave function,
represents the extremely tough challenge in the fields of quantum computing, quantum information, and condensed matter physics. 
In this work, by the example of the well-structured Dicke states we demonstrate that the complexity of these real-valued wave functions can be accurately tuned by introducing a random-sign structure, 
which allows us to explore the regime of the volume-law entanglement. 
Importantly, setting nontrivial sign structure one can increase the entanglement entropy of the Dicke state to the values that are close to Page's estimates for Haar-random states. 
The practical realization of these random-sign Dicke states is possible on different physical platforms with shallow quantum circuits. 
On the level of the measurements the change in the quantum state complexity due to sign structure can be traced out with the dissimilarity measure that estimates multi-scale variety of patterns in bit-string arrays.
\end{abstract}

\maketitle

Complexity is a cornerstone concept for various disciplines~\cite{Anderson1972}, which can be described and understood at the different levels by using various measures~\cite{Lloyd2001}. 
In the context of quantum computing~\cite{Brassard1998,Ladd2010,Fedorov2022}, 
the question of the computational complexity of various problems starts to be linked to the analysis of the structure of corresponding many-body wave functions.
Indeed, certain classes of quantum systems are considered to be efficiently simulated with the use of classical resources, for example, 
tensor networks of the matrix-product-state type are particularly well suited for describing gapped one-dimensional lattice systems with local interactions~\cite{Orus2019}.
One may argue that such systems have sufficiently low level of quantum entanglement, which is considered as the key resource~\cite{Vidal2003,Jozsa2003} for achieving quantum advantage. 
However, not only the amount, but also {\it the structure} of entanglement plays the role since there are known examples of highly-entangled quantum states, 
which can be efficiently simulated with classical computations, for example, quantum systems with area-law entanglement~\cite{Vedral2008,DasSarma2017,Orus2019} (such as aforementioned one-dimensional spin chains with local interaction) and 
Clifford quantum circuits~\cite{Gottesman1998,Gottesman2004,NielsenChuang2000}.
Thus, the complexity of quantum many-body wave functions becomes an important and non-trivial subject for investigations~\cite{Nielsen2005,Aaronson2016,Jefferson2017,Aspuru-Guzik2021}. 
A special attention in this direction is paid to the case of quantum states and quantum circuits, which has a tunable parameter allowing transition between various complexity regimes. 
In other words, the idea is that one can start with a setting that is classically simulatable and tune a specific parameter to increase the complexity. 
A similar idea is explored in the experiment on quantum computational advantage with random circuits~\cite{Morvan2023}, where at the certain point
the computational cost of the experiment achieves the level that is beyond the capabilities of existing classical supercomputers even in the presence of noise. 
In addition, the recent analysis~\cite{Gorshkov2023} has demonstrated a possibility to pass through a complex phase transition generated by entanglement in the case of
the task of simulating single-qubit measurements of $k$-regular graph states on $n$ qubits.

One of the supplementary approaches to control the structure of entanglement in quantum systems is related to the intricate sign structure of the corresponding wave function~\cite{Grover2015}.
Specifically, a nontrivial sign structure prevents one from efficient simulating frustrated quantum magnets described with the antiferromagnetic Heisenberg models on triangular or kagome lattices 
by using quantum Monte Carlo and new variational neural quantum state approaches~\cite{Bagrov2020,Bagrov2023}. 
At the same time, the sign structure of a quantum state could be very sensitive to the basis choice so that for some instances~\cite{Marshall1955} 
it becomes possible to get the non-negative wave function with appropriate unitary transformations~\cite{Fisher2020}, which is exponentially hard to perform in the general case. 
Therefore, one of the ways to study transition between different classes of simulatable and non-simulatable quantum states is to control the sign structure. 

\begin{figure}[!th]
	\includegraphics[width=\linewidth]{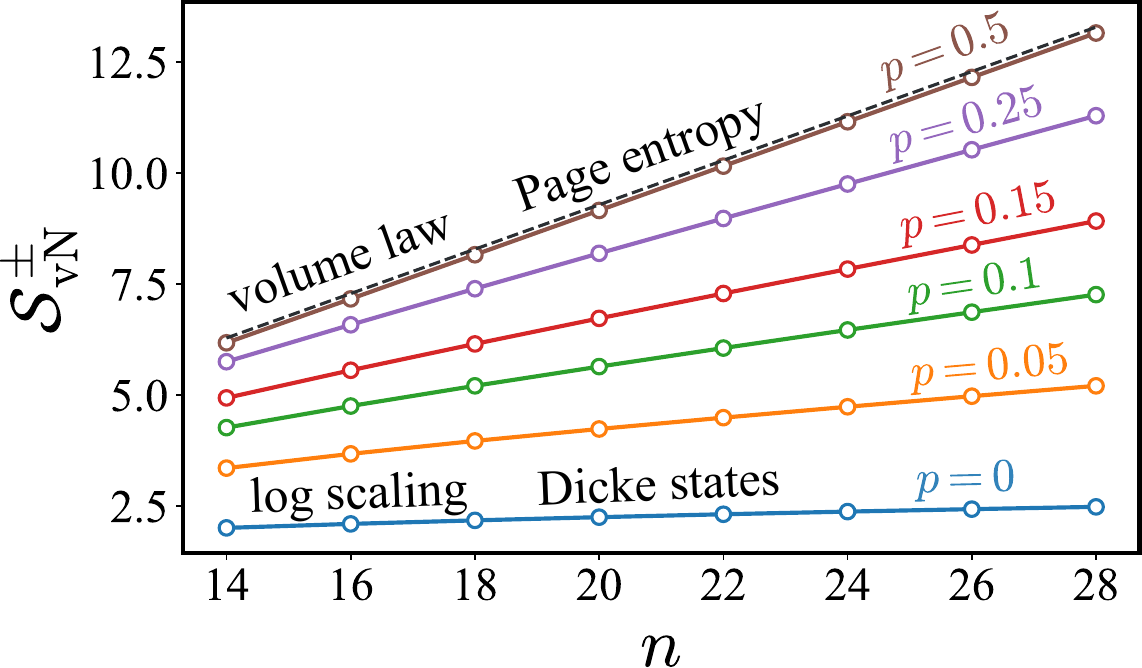}
	\caption{\label{entropy} Comparison of the dependencies of the von Neumann entropies calculated for ordinary Dicke states (blue line) and random-sign Dicke states on the system size. 
	Dashed line denotes entropy of Haar-random states. 
	In all the cases we consider half-filled wave functions with $k = n/2$.  The presented dependencies were averaged over 128 instances. The corresponding standard deviation is smaller than the symbol size.}
\end{figure}

In this work, we demonstrate that manipulating sign structure of rather simple wave functions opens a way to strengthen quantum correlations in system in question and explore the transition from logarithmic growth to the volume-law entanglement. 
A promising candidate is the well-known family of the Dicke states~\cite{Dicke1954,Yudson1985,Kurlov2023}
\begin{eqnarray}\label{Dicke_wf}
	\Ket{D^{k}_{n}} = \frac{1}{\sqrt{C^k_n}} \sum_{j} P_{j}(\Ket{0}^{\otimes n-k} \otimes \Ket{1}^{\otimes k}),
\end{eqnarray}
which represent a paradigmatic model of quantum optics and describe $k$ excitations in $n$-qubit systems; the sum goes over all possible permutations $P_j$ of qubits, and $C_n^k$ is the number of $k$-combinations from a set of $n$ elements.
We note that Dicke states and related quantum-optical models have been intensively studied with the use of 
quantum technologies and quantum simulation~\cite{Wallraff2008,Gross2008,Zeilinger2009,Brennecke2013,Deutsch2017,Chen2017,Ustinov2017,Nori2019,Sedov2020,Ricco2022,Luchnikov2022,Jia2024,ourBS}. 
These well-structured states with positive amplitudes characterized by logarithmic scaling of the entanglement have interesting applications in quantum networking protocols \cite{Zeilinger2009,Jia2024}, 
decoherence-protected quantum codes~\cite{Ouyang2014} and considered as a standard benchmark in quantum states tomography~\cite{Torlai2018,Tiunov2020,Passetti2023,Lvovsky2023} or certification tasks~\cite{Sotnikov2022}. 
As we show, the Dicke wave function with the imposed random sign structure can be characterized by extreme values of the entanglement entropy typical for quantum states picked at random in the full Hilbert space. 
We also propose a protocol based on constructing compact quantum circuits, which can be used for direct experimental realization of such random sign wave functions.
In order to confirm the amplification of the entanglement entropy due to the sign structure on the level of measurements, one can use the dissimilarity measure~\cite{Sotnikov2022} for analysis of the bit-string arrays.        

{\it Sign structure.} ---  We introduce the random-sign Dicke state $\Ket{\pm D^{k}_{n}}$, obtained from $\Ket{ D^{k}_{n}}$ by independent random flipping of a phase of each involved basis state performed with flip probability $p$.
The resulting state can be written as follows:
\begin{eqnarray}\label{Dicke_wf_sign}
	\Ket{\pm D^{k}_{n}} = \frac{1}{\sqrt{C^k_n}} \sum_{j} s_j P_{j}(\Ket{0}^{\otimes n-k} \otimes \Ket{1}^{\otimes k}),
\end{eqnarray}
where $s_j=\pm 1$ denotes the randomly chosen sign for $j$th permutation ($\Pr[s_j=-1]=p$).

To explore the influence of the random sign structure on the properties of the Dicke wave functions we first consider the small-size quantum systems of 14, 16, 18, 20, 22, 24, 26 and 28 qubits. For each system size and the given probability $p$ we generate 128 random sign Dicke states.
To characterize thus generated quantum states we use the von Neumann entanglement entropy, $\mathcal{S}^{\pm}_{\rm vN} (\rho_{A}) = - {\rm Tr} \rho_{A} \log_2 \rho_{A}$, 
where $\rho_{A} = {\rm Tr}_{B} {\Ket{\pm D^{k}_{n}} \Bra{\pm D^{k}_{n}}}$ is the reduced density matrix for the subsystem $A$ which is the half of the system in question (${\rm Tr}_{B}$ stands for the partial trace over the remaining half $B$). 

Figure~\ref{entropy} shows the dependence of the calculated von Neumann entropy of the Dicke states with the perfectly balanced bipartition $k = n/2$ on the value of $p$. For $p=0$ which is the case of the ordinary Dicke wave functions, $\mathcal{S}_{\rm vN} (\rho_{A})$ demonstrates logarithmic dependence on the system size. The calculated von Neumann entropy agrees with analytical results reported in Refs.\cite{Dicke_entropy1, Dicke_entropy2}, where the approximate expression $\mathcal{S}_{\rm vN} (\rho_{A}) \approx a \log_2 \frac{n}{2} + b$ was likewise proposed. Here the fitting parameters $a$ and $b$ can be slightly varied to get the best agreement depending on the scale of the considered systems. In our case they are $a = 0.435$ and $b = 0.787$. 

The presence of a small fraction of the basis states with negative amplitudes considerably affects the complexity of the Dicke state, $\mathcal{S}^{\pm}_{\rm vN} (\rho_{A})$ grows much faster than that for $p = 0$. Remarkably, at the probability $p = 0.5$ the calculated entropy of the $\Ket{\pm D^{k}_{n}}$ wave function is proportional to $n$ and can be accurately fitted for the considered systems and range of qubit numbers with the following expression: 
\begin{eqnarray}
	\mathcal{S}^{\pm}_{\rm vN} (\rho_{A}) = \frac{n}{2} - \frac{4}{5}, 
\end{eqnarray}
where ${n}/{2}$ is the number of qubits in the subsystem $A$.
Such a dependence of the von Neumann entropy on the system size observed for the random sign Dicke functions is similar to the behaviour of the Haar-random state characterized by the average entanglement entropy  $\mathcal{S}^{\rm Page}_{\rm vN} (\rho_{A}) = \frac{n}{2} - \frac{1}{2 \ln 2}$ \cite{Page1993}, which is close to the maximal entropy value $\mathcal{S}^{\rm max}_{\rm vN} (\rho_{A}) = \frac{n}{2}$ for the subsystem A. In the Supplementary Materials \cite{supp} we show that the results obtained for randomly chosen bipartitions with both the same and different number of qubits in the subsystems also confirm the existence of the volume-law transition for the random-sign Dicke states generated at the probability of 0.5.
One should note that in contrast to Dicke states with well-defined structure, the random states are completely delocalized in the Hilbert space and obey to the Porter-Thomas law \cite{Martinis2019,Morvan2023} 
with respect to the probabilities of the basis states. 

As the simplest example that demonstrates enhancement of the entropy and allows analytical consideration we discuss 
$\Ket{D^{2}_{4}}$ and $\Ket{\pm D^{2}_{4}}$ wave functions, which are given by
\begin{eqnarray}
\label{D24}
	\frac{\Ket{0011} + \Ket{0101} + \Ket{0110} + \Ket{1001} + \Ket{1010} + \Ket{1100}}{\sqrt{6}}  
\end{eqnarray}
and
\begin{eqnarray}
\label{pmD24}
	\frac{\Ket{0011} + \Ket{0101} - \Ket{0110} + \Ket{1001} + \Ket{1010} + \Ket{1100}}{\sqrt{6}},  
\end{eqnarray}
respectively. 
The former non-negative wave function can be rewritten by using the Schmidt representation based on the singular value decomposition as follows \cite{NielsenChuang2000}:
\begin{eqnarray}
\label{tensor_product}
\Ket{D^{2}_{4}} = \sum_{i=1}^{4} \lambda_i \Ket{u_i}_{A} \otimes \Ket{v_i}_{B}, 
\end{eqnarray}
where $\lambda_1 = \sqrt{\frac{2}{3}}$, $\lambda_2 = \lambda_3 = \frac{1}{\sqrt{6}}$, $\lambda_4 = 0$ and the orthonormal vector sets for subsystems $A$ and $B$ are $\{\Ket{u_i}_{A}\}_i = \{ \frac{1}{\sqrt{2}} (\Ket{01}_A + \Ket{10}_A), \Ket{00}_A, \Ket{11}_A, \frac{1}{\sqrt{2}} (\Ket{01}_A - \Ket{10}_A) \}$ and $\{\Ket{v_i}_{B}\}_i = \{ \frac{1}{\sqrt{2}} (\Ket{01}_B + \Ket{10}_B), \Ket{11}_B, \Ket{00}_B, \frac{1}{\sqrt{2}} (\Ket{01}_B - \Ket{10}_B) \}$, respectively. In this case the von Neumann entropy for the subsystem $A$ is defined as $\mathcal{S}_{\rm vN} (\rho_A) =  - \sum_{i = 1}^{4} \lambda^2_i \log_2 \lambda^2_i \approx 1.252$. 

In turn, Schmidt decomposition of the $\Ket{\pm D^{2}_{4}}$ state is characterized by the rank of 4, $\lambda_1 = \lambda_2 = \frac{1}{\sqrt{3}}$,  $\lambda_3 = \lambda_4 = \frac{1}{\sqrt{6}}$. The corresponding orthonormal vector sets for subsystem $A$ and $B$ are defined as $\{\Ket{\pm u_i}_{A}\}_i = \{ \frac{1}{\sqrt{2}} (\Ket{01}_A - \Ket{10}_A), \frac{1}{\sqrt{2}} (\Ket{01}_A + \Ket{10}_A), \Ket{11}_A, \Ket{00}_A  \}$ and $\{\Ket{\pm v_i}_{B}\}_i = \{ -\Ket{10}_B, \Ket{01}_B, \Ket{00}_B, \Ket{11}_B \}$, respectively. In contrast to $\Ket{D^{2}_{4}}$ the basis set describing subsystem $B$ in the Schmidt decomposition of the Dicke with sign structure contains only trivial wave functions. Thus, modifying the quantum state with the negative amplitude for one of the basis states results in enhancement of the entanglement entropy from 1.252 to 1.918.

\begin{figure}[!b]
	\includegraphics[width=0.99\linewidth]{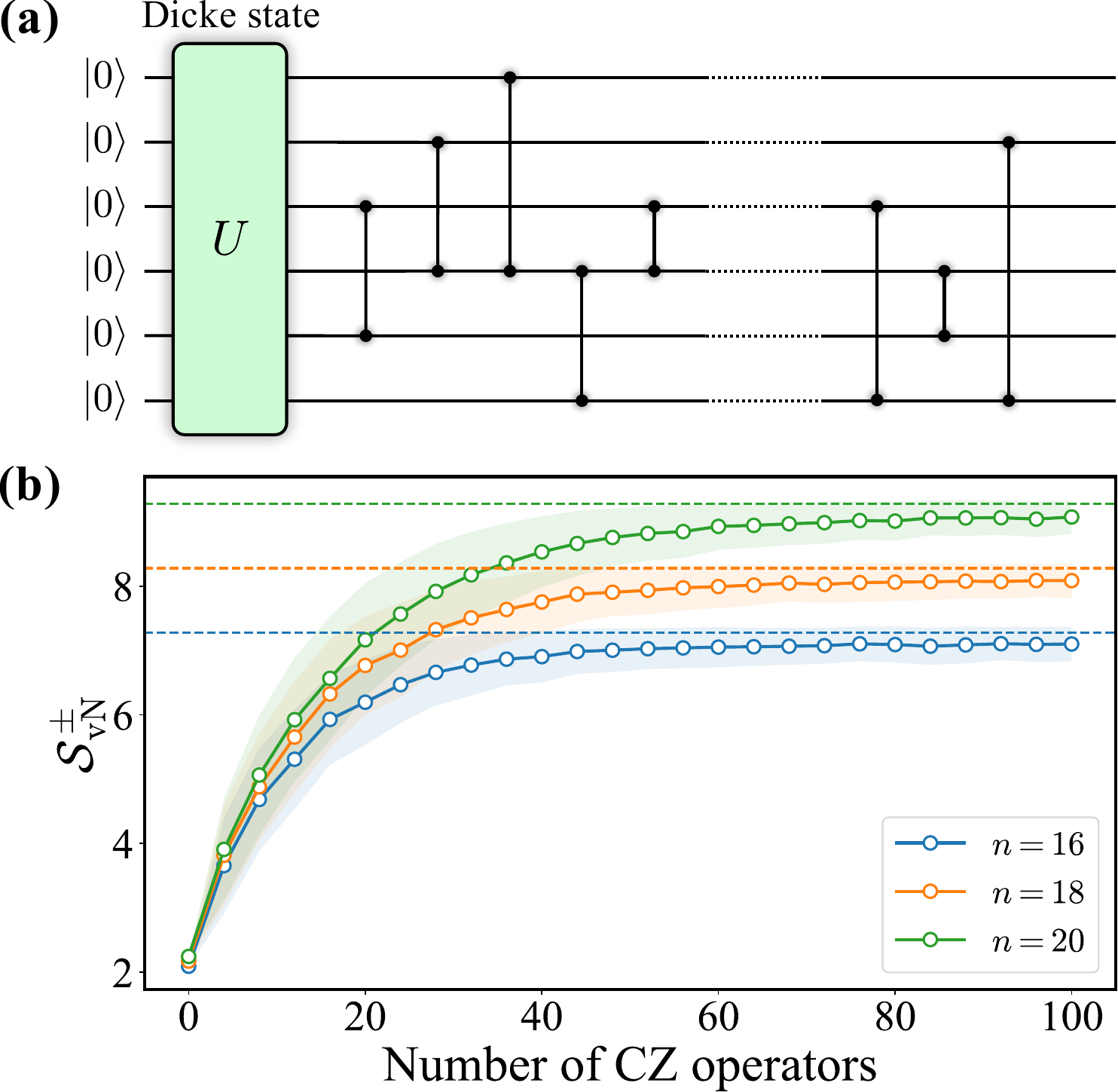}
	\caption{\label{quantum_circuit} (a) Schematic representation of the circuit to initialize Dicke state characterized by a random sign structure. 
	First, $\Ket{D^{k}_{n}}$ is prepared according to the procedure described in Ref.\cite{Eidenbenz2019}. 
	Then, the random sign structure of the wave function is provided with a number of the CZ gates in randomly chosen pairs of qubits. 
	(b) Calculated von Neumann entropy of random-sign Dicke states with $k = \frac{n}{2}$, generated with the protocol for $q = 0.5$. The data were averaged over 128 instances and 10 different bipartitions. Dashed lines denote the Page's entropy $\mathcal{S}_\text{vN}^\text{Page}$ for each system size $n$.}
\end{figure}

{\it Quantum circuits for studying the Dicke states with sign structure} --- From the perspective of the implementation of the Dicke state (and related models) in real experiments, 
one can employ various protocols developed for different quantum platforms including superconducting qubits \cite{Wallraff2008,Gross2008,Chen2017,Ustinov2017}, trapped ions \cite{Wineland2009}, and atomic ensembles~\cite{Mabuchi2004}.  
In this regard, important results have been reported in Ref.~\cite{Eidenbenz2019}, 
where an efficient scheme to initialize real-valued Dicke wave functions with positive amplitudes by means of shallow quantum circuits has been proposed. 
On this basis one can impose nontrivial sign structure and, as it was shown in the previous section, increase the complexity of the system in question. In general, to change a sign of the specific basis function one is to use the multi-qubit controlled-Z (CC...CZ) gate that encodes the corresponding basis state. However, the number of such multi-qubit gates that is required for achieving the volume-law regime ($p = 0.5$) should be equal to the half of the state space dimension of the corresponding Dicke wave function, which is unrealistically large. An additional complexity arises from the fact that for constructing quantum circuits on real quantum devices one needs to transpile each multi-qubit controlled-Z gate with several dozen one- and two-qubit gates \cite{transpileCCZ1, transpileCCZ2}. This obviously requires finding alternative ways to impose sign structures on the level of quantum circuits, which is one of the main goals of this research.

The scheme for creating $\Ket{\pm D^{k}_{n}}$ states proposed by us is presented in Fig.~\ref{quantum_circuit}\,(a). 
It contains two principal parts. 
Starting from trivial $\Ket{0000\dots0}$ state one first prepares the ordinary Dicke wave function, $\Ket{D^{k}_{n}}$  with $U^{k}_{n}$ operator described in Ref.~\cite{Eidenbenz2019}.
The detailed description of such an operator and the corresponding gates for the case of the 4-qubit system are given in \cite{supp}. 
The second part of the quantum circuit consists of the array of CZ gates randomly distributed across the quantum system. 
More practically, for the randomly chosen pair of qubits in the system we add the CZ gate with probability $q$. The total number of such gates with repetitions required to achieve the volume-law regime does not exceed $\frac{n^2}{2}$.

The performance of the proposed scheme can be estimated by the 16-, 18- and 20-qubit examples obtained with $q=0.5$ presented in Fig.~\ref{quantum_circuit}\,(b). In contrast to the random-sign Dicke states obtained with Eq.\eqref{Dicke_wf_sign} the quantum circuit protocol produces the wave functions characterized by the finite dispersion of the entanglement entropy, which is due to the approximation of the CC...CZ gates with CZ ones. Nevertheless, in all the cases we observe saturation of the entanglement entropy to the values that are close to Page's estimates for the random states.

\begin{figure}[!t]
	\includegraphics[width=\linewidth]{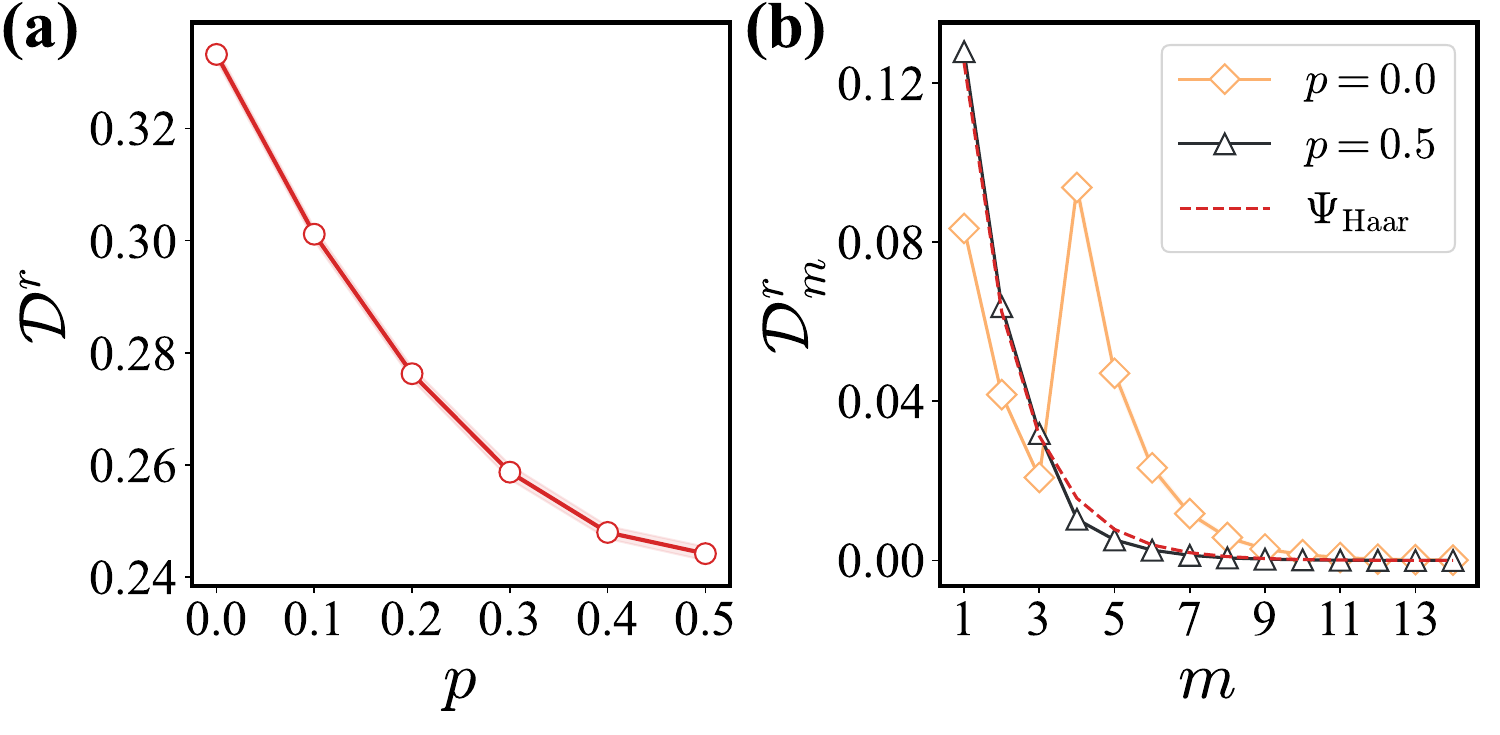}
	\caption{\label{dissimilarity_signed_dicke_16} (a) Full and (b) partial dissimilarity of bits measured in random basis for 16-site Dicke states with sign structure. Data is averaged over 100 states, each state was measured 8192 times. The corresponding standard deviation is smaller than the symbol size.}
\end{figure}

{\it Probing the entanglement entropy.} --- The possibility to manipulate the complexity of the Dicke state by changing its sign structure demonstrated above by calculating the von Neumann entropy should be likewise confirmed on the level of the measurements. 
For these purposes, we employ the structural complexity or dissimilarity measure proposed by some of us in Refs.~\cite{Mazurenko2020,Sotnikov2022}. 
Generally, this quantity estimates the variety of patterns or structures at different spacial scales of an object~\cite{Wolpert_dissimilarity}, which is a bit-string array in our case. 
It practically means that one performs a multi-step renormalization of patterns and calculates the overlap between neighboring renormalized layers.
At each step of the renormalization procedure, the whole bit-string array is divided into non-overlapping windows (filters) of the same size of $\Lambda^m$, where $m$ is the number of the renormalization step.  
Such a procedure allows certifying quantum states of different complexity~\cite{Sotnikov2022, Mazurenko_DMI}, explore classical and quantum systems out of equilibrium~\cite{Mazurenko2024}, 
detect quantum phase transitions \cite{Khatami2022,Khatami2023}, etc.

According to Ref.~\cite{Sotnikov2022} the certification of the positive-amplitude Dicke states can be done by calculating the dissimilarity in two different measurement bases, 
the standard $\sigma^z$ (dissimilarity $\mathcal{D}^z$) and random basis (dissimilarity $\mathcal{D}^r$). Importantly, these quantities are sensitive to the change of the entanglement entropy of the state in question. 
Thus, they can be used for characterizing the quantum state in situations when the direct calculations of $\mathcal{S}_{\rm vN}$ are not possible, which are the cases of performing real experiments or simulating large-scale quantum systems. 
The corresponding methodological details are presented in \cite{supp}. 

Since $\Ket{D^{k}_{n}}$ and $\Ket{\pm D^{k}_{n}}$ are the same from the point of view of projective measurements in the $\sigma^z$ basis, we concentrate on the analysis of the random basis dissimilarity, $\mathcal{D}^{r}$. 
Fig.~\ref{dissimilarity_signed_dicke_16}\,(a) demonstrates a gradual decrease of the random basis dissimilarity as the probability of imposing negative sign increases. 
These results clearly justify the possibility to discriminate the wave functions without ($\mathcal{D}^{r}$ = 0.33) at $p=0$ and with ($\mathcal{D}^{r}$ = 0.245) sign structure at $p=0.5$.  
Remarkably, the minimal value of the dissimilarity calculated for $\Ket{D^8_{16}}$ is equal to 0.245, which is close to the $\mathcal{D}^{r}$ = 0.25 estimated in Ref.~\cite{Sotnikov2022} for Haar-random states of the same number of qubits. 

Proximity of the random sign Dicke state to the Haar-random state can be further elaborated with partial dissimilarity [Fig.~\ref{dissimilarity_signed_dicke_16}\,(b)]. 
One can see that the dissimilarity of the random sign Dicke state as a function of the renormalization step, $m$ slightly deviates from the simple analytical law, 
\begin{eqnarray}
	\mathcal{D}_{m} = \frac{1}{2}(1- \Lambda^{-1})\Lambda^m
\end{eqnarray}
derived for the Haar-random states in Ref.~\cite{Sotnikov2022}.
It means that the resulting dissimilarity only depends on the filter size, $\Lambda$ and is insensitive to the content of the bit-string array. 
In turn, the non-negative wave function reveals completely different partial dissimilarity profile that is characterized by two maximal values at $m=1$ and $m=4$. Position of the second peak depends on the system size and occurs when one starts taking into account patterns which are larger in scale than a single bitstring. Importantly, the $\mathcal{D}^{r}_4$ value is sensitive to the diversity of the presented bitstrings in the considered array of measurements and decreases when increasing $p$. In the Supplementary Materials \cite{supp} we discuss this issue in more details as well as demonstrate that calculating dissimilarity allows distinguishing ordinary Dicke wave functions and those with sign structure even in the presence of noise and gates imperfections that are inevitable for real quantum devices. Thus, the formation of the random sign Dicke states can be unambiguously confirmed by analyzing the measurement results with the dissimilarity measure.

{\it A parent Hamiltonian for the Dicke states with sign structure.} ---
It is important to discuss constructing effective Hamiltonians whose ground states are the introduced Dicke states with imposed sign structure. One of the possibilities to realize the ordinal Dicke states, $\Ket{D^k_n}$ in real experiments is related to the Lipkin-Meshkov-Glick Hamiltonian \cite{LMG} with anisotropic in-plane ferromagnetic couplings and z-oriented external magnetic field term. The main peculiarity of the LMG  Hamiltonian is that the interaction strength is the same for all the pairs of spins in the system, which means that the model is defined on the complete graph. Nevertheless, the LMG Hamiltonian can be realized experimentally \cite{LMG_experiment}.  

Here we consider a simplified version of the LMG model, where one neglects in-plane anisotropy of the exchange interactions,
\begin{eqnarray}
\label{LMG_Ham}
H_{\rm LMG} = - \sum_{i < j} (\sigma_{i}^x \sigma_{j}^x + \sigma_{i}^{y} \sigma_{j}^{y}) - h \sum_{i} \sigma_{i}^{z},
\end{eqnarray}
where $\sigma_{i}^{\mu}$ ($\mu = x,y,z$) is the Pauli matrix. 
At $h=0$ the ground state of the Hamiltonian with even number of spins $N$ is nothing but the Dicke state $\Ket{D^k_n}$ with $k=\frac{n}{2}$. Switching the magnetic field on leads to a change of the Dicke index of the ground state wave function. The critical value of the magnetic field that corresponds to the transition from $\Ket{D^{k}_n}$ to $\Ket{D^{k+1}_n}$ state
is given by $h_{\text{crit}} = n - 2k - 1$.

Starting with the Hamiltonian for the LMG model that satisfies $H_{\rm LMG} \Ket{D^k_n} = E^{k}_{n} \Ket{D^k_n}$ we are to construct ${\hat H}_{\pm}$ that is characterized by the same eigenspectrum, $E^{k}_{n}$ and eigenfunctions $\Ket{\pm D^k_n}$ that are the Dicke states with sign structure. For that we first express the CZ gate between neighbouring ith and jth qubits through the Pauli matrices as 
\begin{eqnarray}
{\rm CZ} (i,j) = \frac{I_i \otimes I_j + \sigma^z_i \otimes I_j + I_i \otimes \sigma^z_{j} - \sigma^z_{i} \otimes \sigma^z_{j}}{2},
\end{eqnarray}
where $I_i$ is identity operator acting on the ith qubit. Then 
all the CZ gates are merged into single $\xi$ operator that allows us to  define the required Hamiltonian as ${\hat H}_{\pm} = {\hat \xi} {\hat H}_{\rm LMG} \hat \xi^{\dagger}$. This demonstrates the main complexity for realizing the random-sign Dicke states with a physical system rather than quantum circuits on a quantum computer. Namely, the corresponding Hamiltonian, $H_{\pm}$ is characterized by multi-spin interactions whose orders depend on the number of the CZ gates required for imposing the particular sign structure.  

For the 4-qubit example with the $\Ket{\pm D^2_{4}}$ wave function (Eq.\ref{pmD24}) characterized by the minus sign for the $\Ket{0110}$ basis function that can be imposed with single CZ gate as discussed in  \cite{supp}. Thus, for the $\Ket{\pm D^2_{4}}$ state we obtain the following parent Hamiltonian
\begin{eqnarray}
{\hat H}_{\pm} = - (\sigma_{0}^{x} \sigma_{1}^{x} + \sigma_{0}^{y} \sigma_{1}^{y}) \sigma^z_{2} - (\sigma_{0}^{x} \sigma_{2}^{x} + \sigma_{0}^{y} \sigma_{2}^{y}) \sigma^z_{1}  \nonumber \\
- (\sigma_{1}^{x} \sigma_{3}^{x} + \sigma_{1}^{y} \sigma_{3}^{y}) \sigma^z_{2} - (\sigma_{2}^{x} \sigma_{3}^{x} + \sigma_{2}^{y} \sigma_{3}^{y}) \sigma^z_{1} \nonumber \\
- (\sigma_{0}^{x} \sigma_{3}^{x} + \sigma_{0}^{y} \sigma_{3}^{y}) - (\sigma_{1}^{x} \sigma_{2}^{x} + \sigma_{1}^{y} \sigma_{2}^{y}). 
\end{eqnarray}
In turn, the physical realization of 6-qubit Dicke states with sign structure requires account of four-spin interactions  \cite{supp}. These results evidence that quantum computer platforms provide the most effective way for creating and manipulating the $\Ket{\pm D^{k}_{n}}$ wave functions introduced in this work.

{\it Conclusions.} --- Further progress in quantum technologies is undoubtedly related to the development of distinct methods for controlling and manipulating entanglement in large-scale quantum systems. 
In this work, we have proposed such an approach that is based on reconstructing the sign structure of real-valued wave functions. 
By the example of the paradigmatically important Dicke states, notable quantum wave functions in quantum optics and computing with storage \cite{storage}, networking \cite{networking}, sensing \cite{sensing} and coding \cite{coding} applications, we demonstrate the amplification of the entanglement by imposing the random-sign structure. 
The upcoming stages of our analysis include optimization of the algorithm for creating sign structure on the number of the two-qubit gates and consideration of alternative methods of preparing the Dicke wave functions~\cite{Cirac1, Cirac2, PRXQuantum.5.010319} as the basis for achieving the volume-law entropy regime.

\begin{acknowledgments}
We thank the support of the Russian Roadmap on Quantum Computing (Contract No. 868-1.3-15/15-2021, October 5, 2021) in the development of the trapped-ion processor.
The work of EOK and AKF is also supported by the Priority 2030 program at the National University of Science and Technology ``MISIS'' under the project K1-2022-027.
\end{acknowledgments}

\section*{Supplementary Materials}
\subsection*{4-qubit example}
In this section we perform the detail characterization of the 4-qubit Dicke states discussed in the main text. 
They are as follows:
\begin{eqnarray}
	\Ket{D^{2}_{4}} = \nonumber \\
\frac{\Ket{0011} + \Ket{0101} + \Ket{0110} + \Ket{1001} + \Ket{1010} + \Ket{1100}}{\sqrt{6}} \nonumber 
\end{eqnarray}
and
\begin{eqnarray}
	\Ket{\pm D^{2}_{4}} = \nonumber \\ \frac{\Ket{0011} + \Ket{0101} - \Ket{0110} + \Ket{1001} + \Ket{1010} + \Ket{1100}}{\sqrt{6}}. \nonumber 
\end{eqnarray}

The corresponding density matrix for Dicke wave function with sign structure is given by 
\begin{equation}
  \label{eq:full_matrix_signed}
  \rho^{\pm} = \Ket{\pm D^{2}_{4}}\Bra{\pm D^{2}_{4}} =  \frac{1}{6}\cdot
  \begin{pmatrix}
    \ \ 1 & \ \ 1 & -1 & \ \ 1 & \ \ 1 & \ \ 1 \\
    \ \ 1 & \ \ 1 & -1 & \ \ 1 & \ \ 1 & \ \ 1 \\
    -1 & -1 & \ \ 1 & -1 & -1 & -1 \\
    \ \ 1 & \ \ 1 & -1 & \ \ 1 & \ \ 1 & \ \ 1 \\
    \ \ 1 & \ \ 1 & -1 & \ \ 1 & \ \ 1 & \ \ 1 \\
    \ \ 1 & \ \ 1 & -1 & \ \ 1 & \ \ 1 & \ \ 1 \\
  \end{pmatrix}. \nonumber 
\end{equation}
Note that here we employ a basis constructed with alphabetically ordered bistrings of Hamming weight 2.
Therefore, the reduced density matrix for subsystem $A$ has the following form in the full two-qubit computational basis
\begin{equation}
  \label{eq:sub_matrix_signed}
  \rho_A^{\pm} = \frac{1}{6}\cdot
  \begin{pmatrix}
    1 & 0 & 0 & 0 \\
    0 & 2 & 0 & 0 \\
    0 & 0 & 2 & 0 \\
    0 & 0 & 0 & 1 \\
  \end{pmatrix}.
\end{equation}
In this case the entropy is defined as follows: 
\begin{equation}
  \label{eq:entropy_sign}
  \renewcommand{\arraystretch}{2}
  \begin{split}
    \mathcal{S}_{\text{vN}}^{\pm} &= \frac{2}{3}\log_2 \,3 + \frac{1}{3}\log_2 \,6 = \log_2 \,3 + \frac{1}{3} \approx 1.918.
  \end{split}
\end{equation}

\begin{figure*}[!t]
	\includegraphics[width=0.9\linewidth]{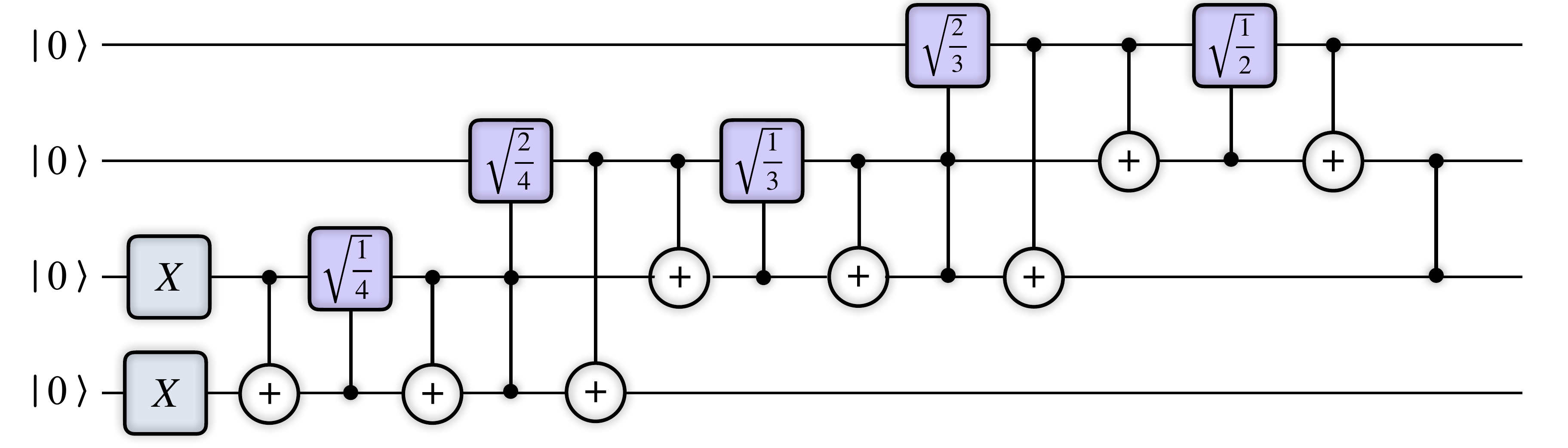}
	\caption{\label{4-qubit_circuit} Representation of the circuit to initialize 4-qubit Dicke state with sign structure, $\Ket{\pm D_{4}^2}$ introduced in the main text. The $\sqrt{\frac{l}{n}}$ gates denote Y-rotation $R_{y}(2\cos^{-1}\sqrt{\frac{l}{n}})$.The sign structure of the wave function is defined by the last CZ gate.}
\end{figure*}

The full density matrix of the non-negative Dicke state is normalized matrix of ones, $\frac{1}{6} \cdot \mathbbm{1}$, which means that the reduced matrix can be written as
\begin{equation}
  \label{eq:sub_matrix}
  \rho_A^{+} = \frac{1}{6}\cdot
  \begin{pmatrix}
    1 & 0 & 0 & 0 \\
    0 & 2 & 2 & 0 \\
    0 & 2 & 2 & 0 \\
    0 & 0 & 0 & 1 \\
  \end{pmatrix}.
\end{equation}

The corresponding eigenvalues of this matrix are $\{\frac{2}{3},\frac{1}{6},\frac{1}{6},0\}$ and the von Neumann entropy is given by 
\begin{equation}
  \label{eq:entropy}
  \renewcommand{\arraystretch}{2}
  \begin{split}
    S_{\text{vN}}^{+} &= \frac{1}{3}\log_2 \,6 - \frac{2}{3}\log_2 \frac{2}{3} = \log_2 \,3 - \frac{1}{3} \approx 1.252.
  \end{split}
\end{equation}

The circuit that prepares the $\Ket{\pm D^{2}_{4}}$ state with sign structure is presented in Fig.~\ref{4-qubit_circuit}. The part of the circuit that contains X gates, CNOT gates and controlled Y-rotation gates was reproduced according to the procedure described in Ref.\cite{Eidenbenz2019}. It allows initializing the non-negative Dicke state $\Ket{D^{2}_{4}}$. In turn, the last controlled-Z gate is responsible for the sign structure.   

\subsection*{Dissimilarity}

In this section we present the details of calculating the dissimilarity of the bit-string arrays, which allows certifying the states with non-trivial sign structure.
Such a procedure originally proposed in Ref.~\cite{Sotnikov2022} includes the processing of the data obtained from the measurements in two different bases, standard $\sigma^z$ basis and a random basis. 

\begin{figure}
	\includegraphics[width=0.9\linewidth]{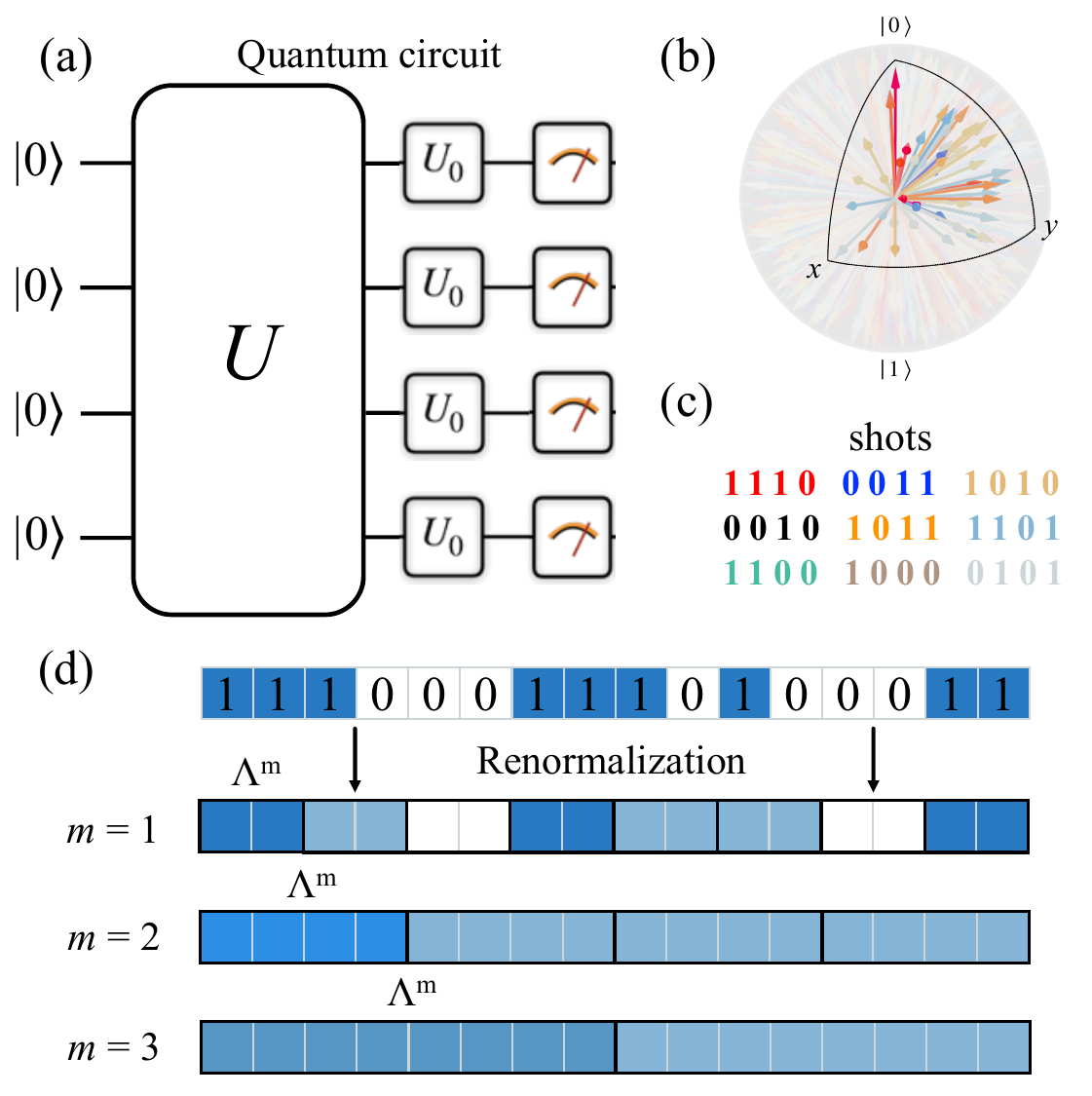}
	\caption{\label{fig:dissim} Schematic representation of the protocol of calculating $\mathcal{D}^{r}$ using the example of a 4-qubit system. (a) After the state is stabilized on a quantum device one uses the universal rotational gates $U_0$ to choose a random point in the specific segment of the Bloch sphere (b). Such a point defines basis for single-shot measurement.  (c) Bitstrings obtained from projective measurements in random basis. (d) First steps of the renormalization procedure. Initial bitstrings are arranged into the one-dimensional array and then coarse-grained several times using the averaging scheme within non-overlapping windows of size $\Lambda^m$. In our calculations we use ``$-1$'' instead of ``0''.}
\end{figure}

To estimate the random-basis dissimilarity $\mathcal{D}^{r}$ of the considered $\Ket{\pm D^{k}_{n}}$ state we initialize such a wave function on the quantum simulator and then perform projective measurements $N_{\rm shots}$ times. Importantly, for each measurement we randomly and independently select an axis to project on from the area of the Bloch sphere shown in Fig.~\ref{fig:dissim}(b). 
More specifically, for each measurement we sample three angles $\theta \in [0, \frac{\pi}{2}]$, $\phi \in [0,\frac{\pi}{2}]$ and $\lambda \in [0, \frac{\pi}{2}]$. 
Then we use these angles as parameters for universal rotational gate, $U_0$ and define a new basis for single measurement for all the qubits. The measurement outputs obtained in such a way are then arranged into one-dimensional sequence of bitstrings $\textbf{b}^0$ of length $L=n\times N_{\rm shots}$. Importantly, to preserve agreement with previously obtained results~\cite{Sotnikov2022} we use ``$-1$'' instead of ``0'' in our calculations.

At every renormalization step $m$, a vector of the same size $L$ is constructed using a simple averaging scheme. 
Its elements are calculated as follows:
\begin{eqnarray}\label{eq:b_i^m}
	b_{i}^{m} = \frac{1}{\Lambda^m} \sum_{l=1}^{\Lambda^m} b_{\Lambda^m[(i-1)/\Lambda^m]+l}^{m-1},
\end{eqnarray}
where square brackets denote taking integer part. This means that the vector $\textbf{b}^{m-1}$ is divided into non-overlapping windows of $\Lambda^m$ size ($\Lambda = 2$ in our calculations),
in which all elements are averaged and substituted with the same value as it is shown in Fig.~\ref{fig:dissim}(d). Index $l$ denotes elements belonging to the same window.

In the simple averaging scheme used in our work dissimilarity between scales $m$ and $m+1$ can be written as follows:
\begin{eqnarray}\label{eq:D_m}
	{\cal D}^r_m = \frac{1}{2L}\left|\left(\sum_{i=1}^L(b_i^{m+1})^2-(b_i^{m})^2\right)\right|.
\end{eqnarray}

Taking into account patterns realized in new scales we obtain  
\begin{eqnarray}
\label{eq:Dissim}
	{\cal D}^r = \sum_{m=1}^M{\cal D}^r_m,
\end{eqnarray}
where $M$ is the total number of renormalization steps.

\begin{figure}[!b]
	\includegraphics[width=\linewidth]{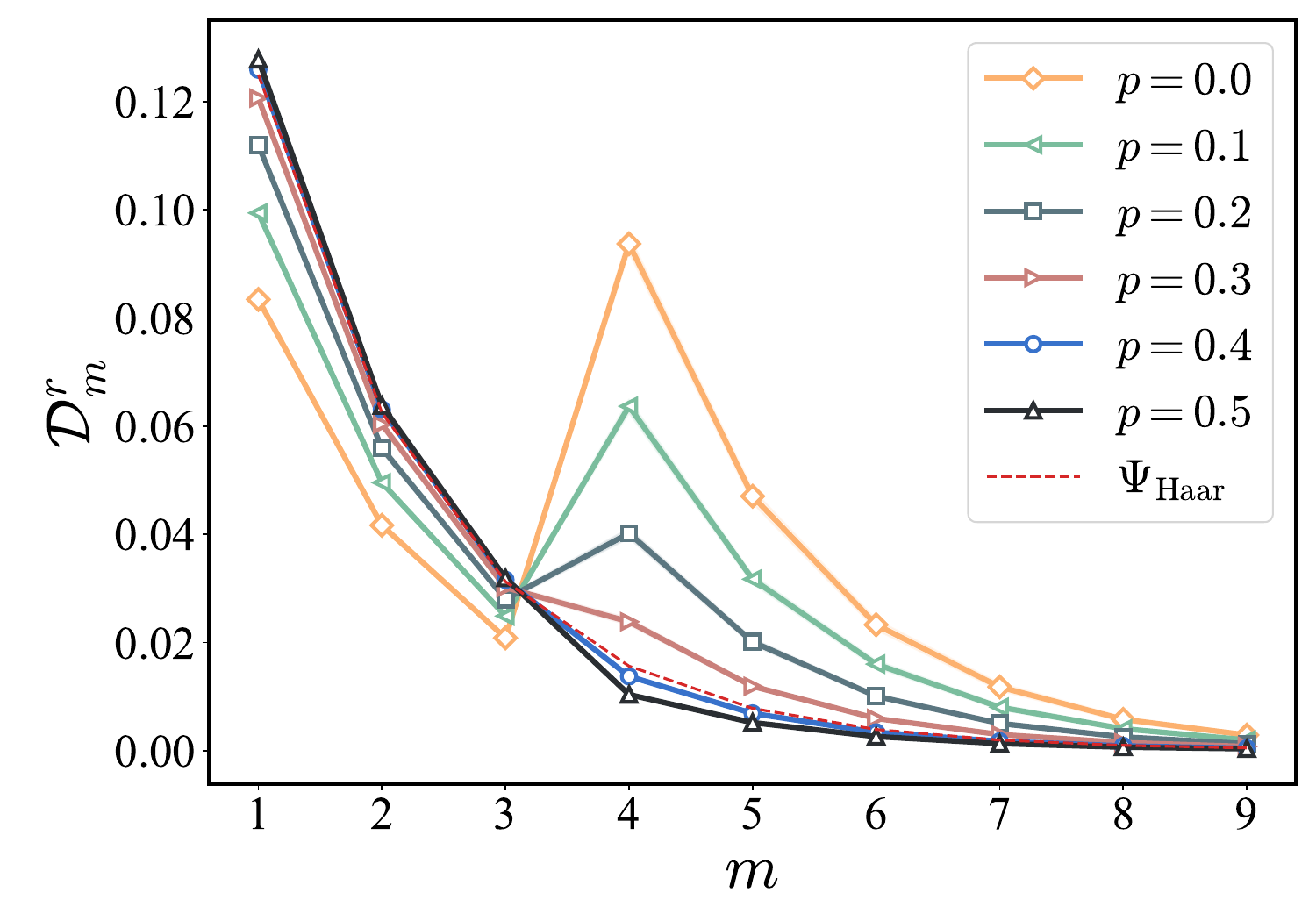}
	\caption{\label{fig:partial_dissimilarity} Partial dissimilarity of bits measured in random basis for 16-site Dicke states with sign structure. Data is averaged over 100 states, each state was measured 8192 times.}
\end{figure}

\begin{figure*}[!t]
  \includegraphics[width=\linewidth]{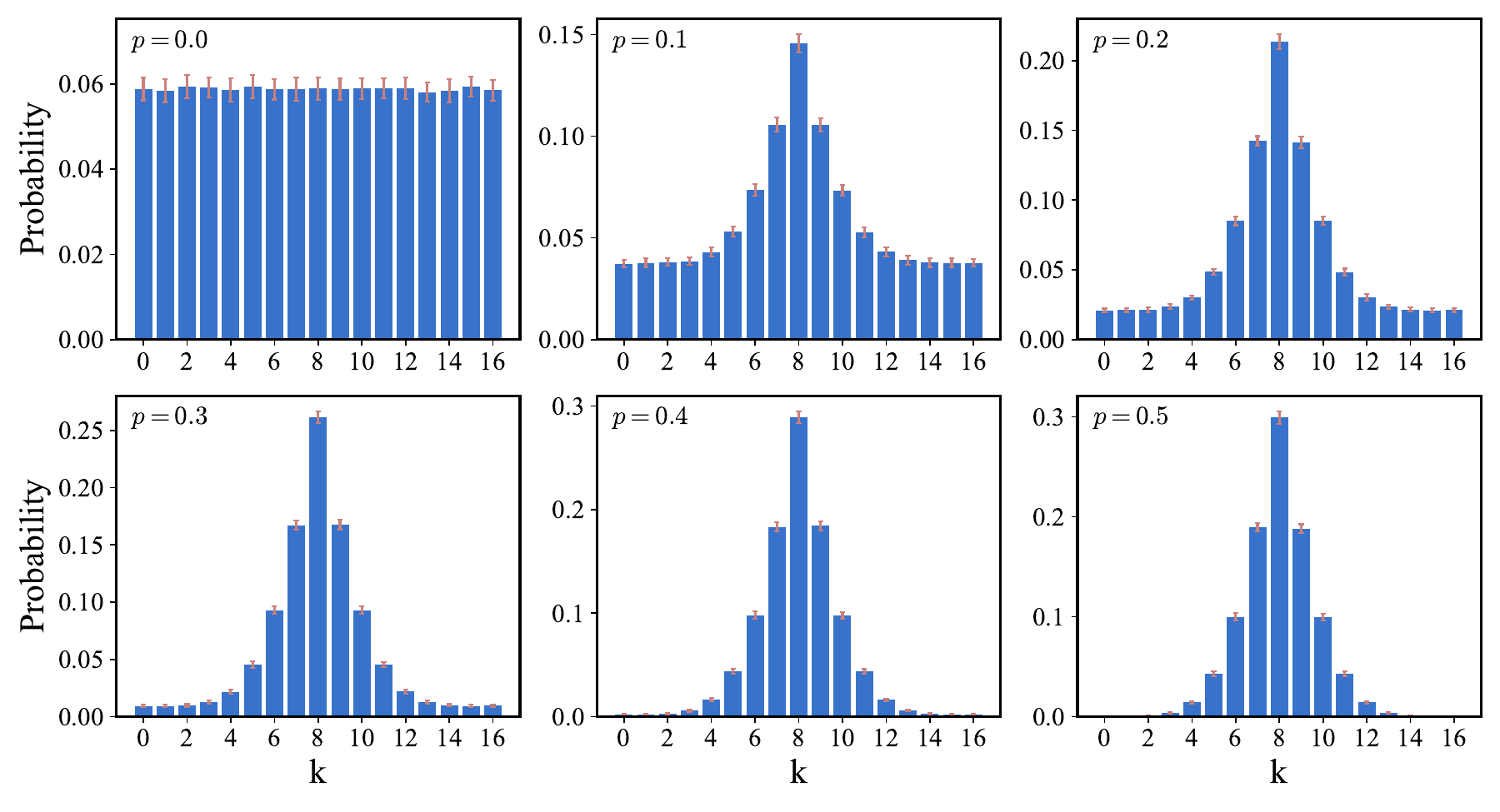}
  \caption{\label{fig:k_distributions} Characterization of the data used for calculating dissimilarity. Distributions of bitstrings measured in random basis over the number of ones in the bitstring for signed structure states with different fraction of minus signs $p$. Each quantum state was measured 8192 times, which means that the ${\bf b}^0$ array contains 8192 bitstrings to calculate the dissimilarity. Data is averaged over 100 states. Red errorbars denote standard deviation of average.}
\end{figure*}

In addition to the dissimilarity results presented in the main text (Fig.3), here in Fig.\ref{fig:partial_dissimilarity} we exemplify the partial dissimilarities obtained from the random-basis measurements for 16-qubit Dicke states with different sign structures. Increasing the complexity of the sign structure of the Dicke state (increasing the probabiity $p$) leads to a gradual suppression of the dissimilarity contribution at the fourth renormalization step. Position of this partial contribution depends on the system size and occurs when one starts taking into account patterns which are larger in scale than the length of the individual bitstring in the ${\bf b}$ array. The behaviour of the $\mathcal{D}^{r}_4$ value with respect to the changes of $p$ fully meets the intuitive expectations and depends on the diversity of the presented bitstrings which is shown in Fig.~\ref{fig:k_distributions}. Namely, the bigger the difference between the elements of ${\bf b}^{m}$ and ${\bf b}^{m+1}$ vectors, the bigger $\mathcal{D}^{r}_m$ one will obtain. 

\begin{figure}
	\includegraphics[width=0.9\linewidth]{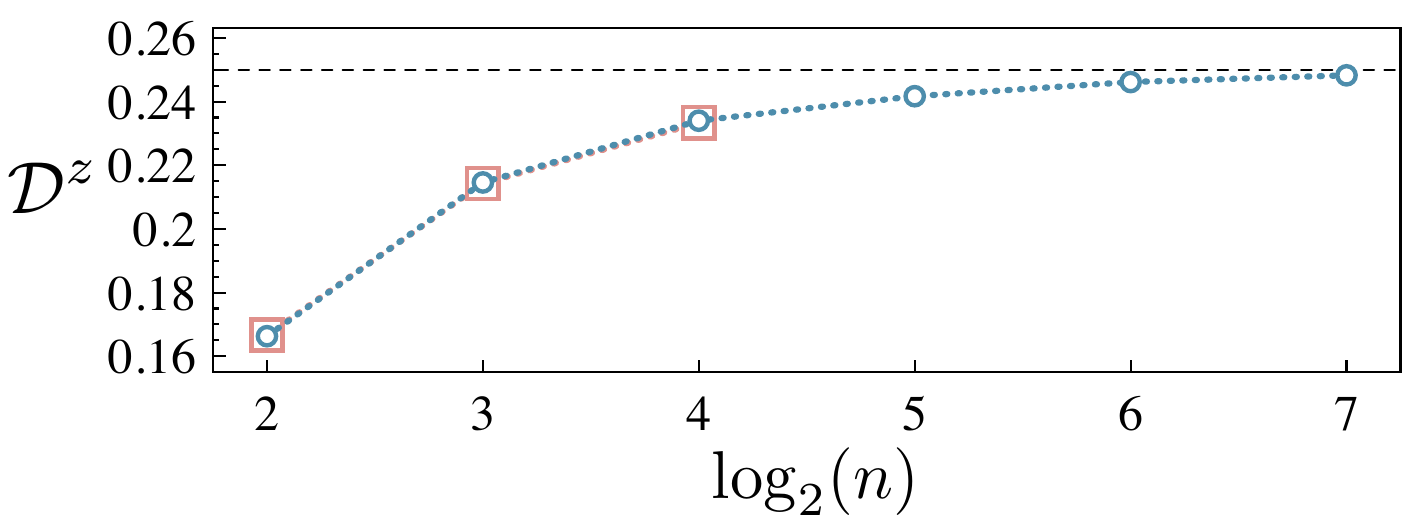}
	\caption{\label{fig:d_z} Dependence of the $\mathcal{D}^{z}$ on the system size obtained for $\Ket{D^{n/2}_{n}}$ states. Red squares correspond to the dissimilarities calculated using the whole basis states while blue circles denote $\mathcal{D}^{z}$ calculated using $2^{13}$ randomly generated bitstrings.}
\end{figure}

It is likewise instructive to evaluate and discuss the dissimilarity in the $\sigma^z$ basis for the systems of different sizes. Since the amount of ``1'''s in each basis state measured in $\sigma_z$ basis is the same, contributions from the steps $m>n$ are equal to zero~\cite{Sotnikov2022}. Therefore, total dissimilarity can be calculated as an averaged ${\cal D}^z$ over all the presented basis states 
\begin{eqnarray}
\label{eq:D_z}
 {\cal D}^z = \sum_{i=1}^{C^k_n}{\cal D}^z_i.
\end{eqnarray}

Considering that the amplitudes of all basis states are the same one can simply randomly choose an appropriate amount of bitstrings ($\ge 2^{13}$) using the uniform distribution to get the ${\cal D}^z$ with a high accuracy. As can be seen from Fig.~\ref{fig:d_z}, the obtained value tends to $0.25$ as the system size increases.

\begin{figure}[!b]
	\includegraphics[width=\linewidth]{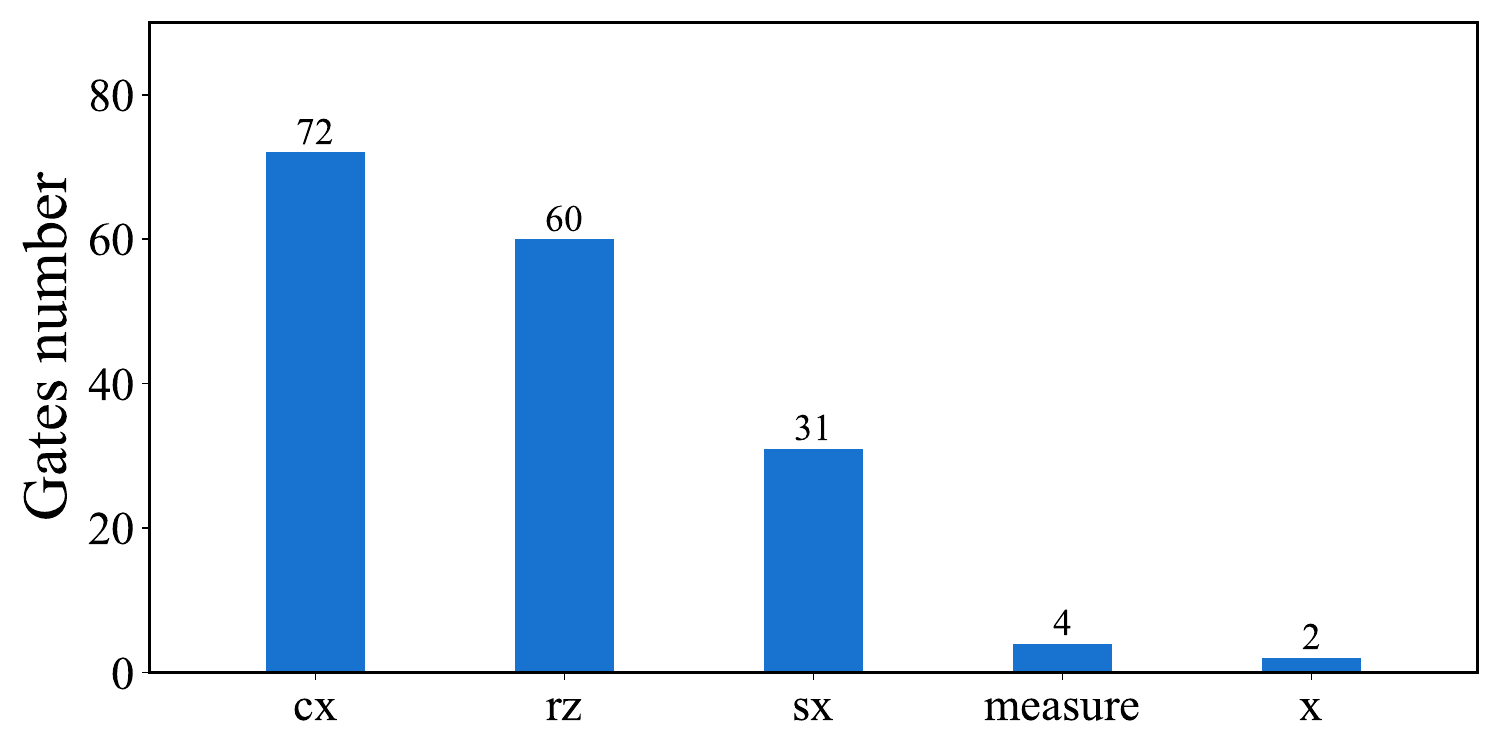}
	\caption{\label{gates_count} Gates number in quantum circuit of the $|\pm{}D_4^2\rangle$ state transpilled to fit the topology of real IBM device (\texttt{ibm\_cairo}).}
\end{figure}

\begin{figure}[!t]
	\includegraphics[width=\linewidth]{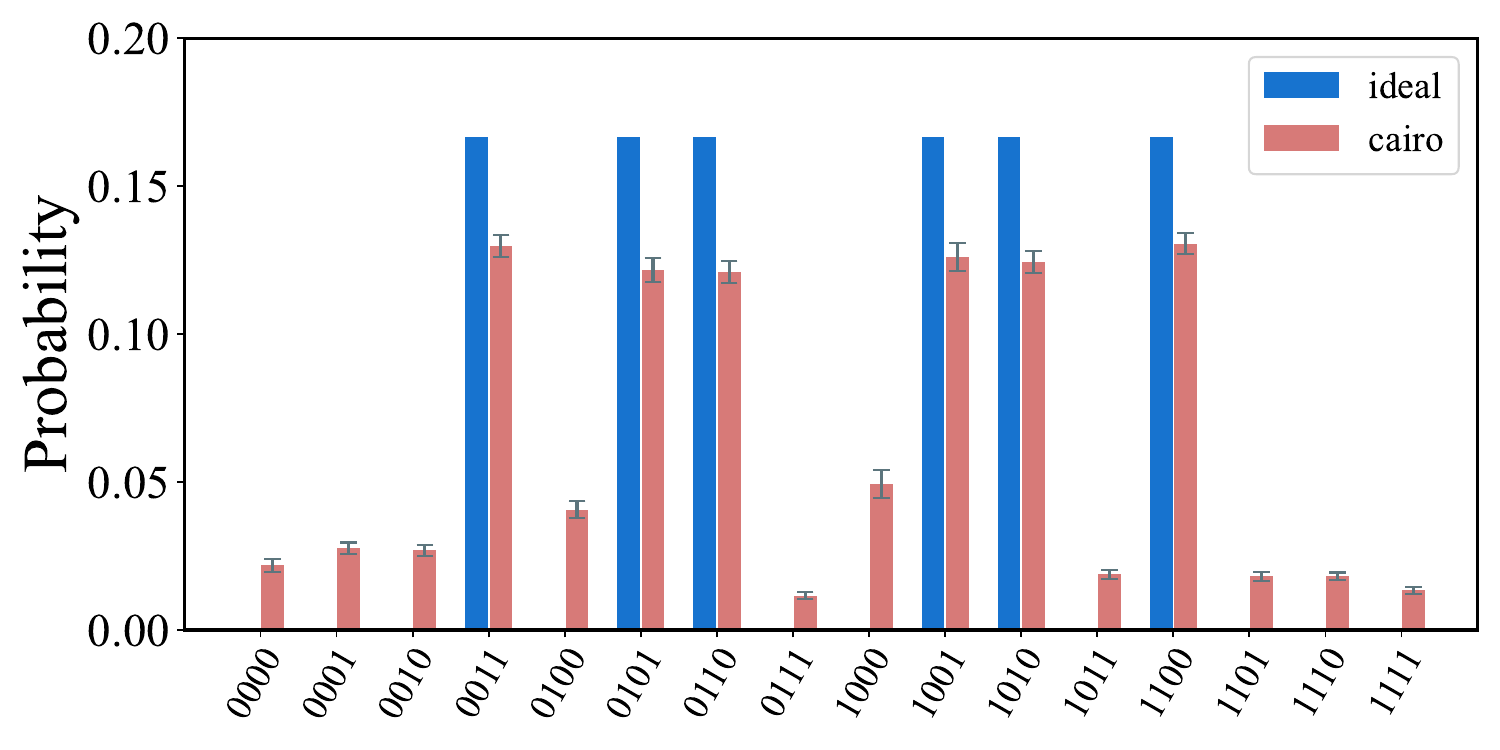}
	\caption{\label{real_state} Comparison of basis states distributions calculated for exact $|D_4^2\rangle$ state (blue color) and obtained from 8192 measurements by using the simulator with noise imitating real \texttt{ibmq\_cairo} device (red color).}
\end{figure}

\subsection{Effect of noise}
The 4-qubit model for Dicke states described above is the minimal one to demonstrate the effect of the sign structure on the entanglement of the wave function. It is important to confirm that our proposals is practical from the point of view of real experiments subjected to decoherence. Put another way, the dissimilarity calculated with noise bitstrings still allows to certify the Dicke states with and without the sign structure. For these purposes, the quantum circuit visualized in Fig.~\ref{4-qubit_circuit} is transpilled to match the topology of the specific quantum device. In addition, we implement the noise model that features the properties of the same device.   

\begin{figure}[!t]
	\includegraphics[width=\linewidth]{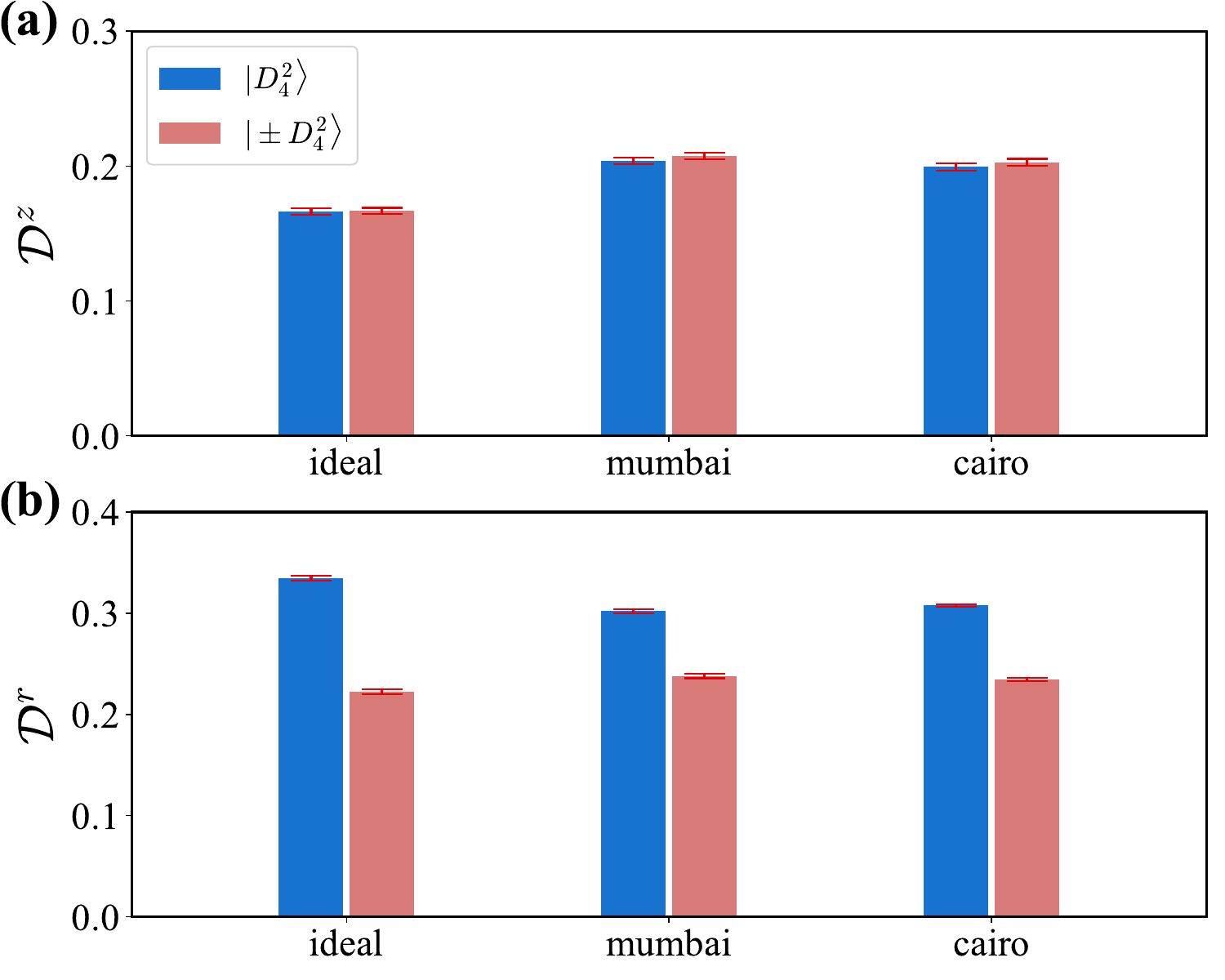}
	\caption{\label{dissimilarity_noise} Dissimilarity of bitstrings measured in (a) z-basis and (b) random basis for $\Ket{D_4^2}$ and $\Ket{\pm{}D_4^2}$ states. The measurements were performed by using ideal quantum simulator and those imitating real IBM devices (\texttt{ibmq\_mumbai} and \texttt{ibm\_cairo}) with corresponding noise models. Each state was measured 8192 times.}
\end{figure}

According to Fig.~\ref{gates_count} the total number of the CNOT gates for preparing $\Ket{\pm{}D_4^2}$ on the real quantum devices provided by IBM is equal to 72. These two-qubit gates are primary interest since their quality or gate error in real devices is mainly responsible for the fidelity of the resulting state. The account of decoherence effects on the level of the noise model with parameters corresponding to the real IBM device considerably affects the target state. Fig.~\ref{real_state} allows estimating the deviation from the ideal $\Ket{D_4^2}$ wave function in terms of the probability distribution of the bitstrings measured in the $\sigma^z$ basis. The obtained distribution still features the largest intensity coming from Dicke-like basis functions, however, all other states likewise produce non-negligible contribution. 

Such a change of the target wave function leads to modification of the dissimilarity we use to characterize the system in question. In the case of the measurements in the $\sigma^{z}$ basis the dissimilarity values, $\mathcal{D}^z$ (Fig.\ref{dissimilarity_noise}) become slightly larger by about 0.04 with respect to the ideal one. In turn, the values of the random-basis dissimilarity calculated for the $\Ket{D_4^2}$ and $\Ket{\pm D_4^2}$ states reveal different trends in changes due to the noise with respect to the noise-free model. Nevertheless, the difference between these states remains statistically robust even in the presence of the gate imperfections.

\subsection{Dependence of the entanglement on the subsystem size}

\begin{figure}[!h]
  \includegraphics[width=\linewidth]{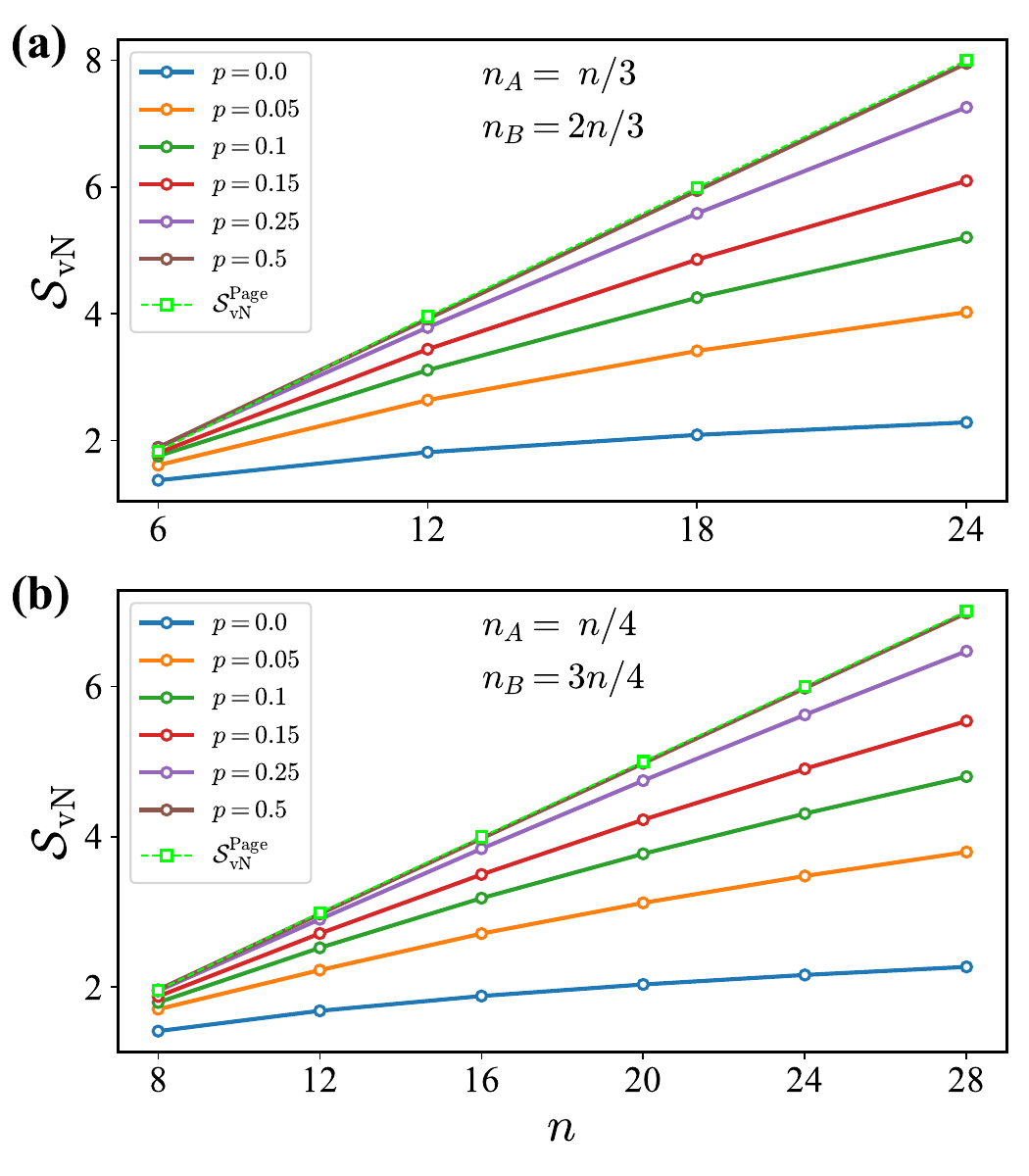}
  \caption{\label{fig:partitioning} Comparison of the dependencies of the von Neumann entropies calculated for ordinary Dicke states (blue line) and random-sign Dicke states on the system size for subsystem sizes of (a) $n/3$ and (b) $n/4$. In all the cases we consider half-filled wave functions with $k = n/2$. The presented dependencies were averaged over 128 instances.}
\end{figure}

The results presented in the main text were obtained for equal bipartition of the system in question, $n_{A} = n_{B}$, where  $n_{A}$ ($n_{B}$) is the number of qubits in the subsystem A (B). In this section we analyze the Dicke state entanglement entropy calculated with different subsystem divisions. Figs.\ref{fig:partitioning} a and b illustrate the cases of $n_{A} = \frac{n}{3}$ ($n_{B} = \frac{2n}{3}$) and $n_{A} = \frac{n}{4}$ ($n_{B} = \frac{3n}{4}$), respectively.
From these data we conclude on the same trend in behaviour of $\mathcal{S}_{\rm vN}$ as one observes for equal bipartition. There is the transition from logarithmic scaling ($p = 0$) to volume law ($p = 0.5$). 

To compare our results with those obtained for random state we use the general formula derived by Page in Ref.\cite{Page1993} for calculating entanglement entropy of the random state at arbitrary bipartition into subsystems A and B that is given by
\begin{eqnarray}
\mathcal{S}^{\rm Page}_{\rm vN} (\rho_{A}) = \frac{1}{\ln 2} \sum_{k=d_{B}+1}^{d_A d_B} \frac{1}{k} - \frac{d_A -1}{2d_{B} \ln 2},
\end{eqnarray}
where $d_{A} = 2^{n_{A}}$ ($d_{B} = 2^{n_{B}}$) stands for the dimension of the subsystem A (B). Importantly, $n_A \le n_B$. The coefficient $\frac{1}{\ln 2}$ is introduced to define the entropy in bits. One can see that the Page's estimates (light green line) coincide with the case of the random-sign Dicke states generated at $p = 0.5$.

\begin{figure}[!h]
  \includegraphics[width=\linewidth]{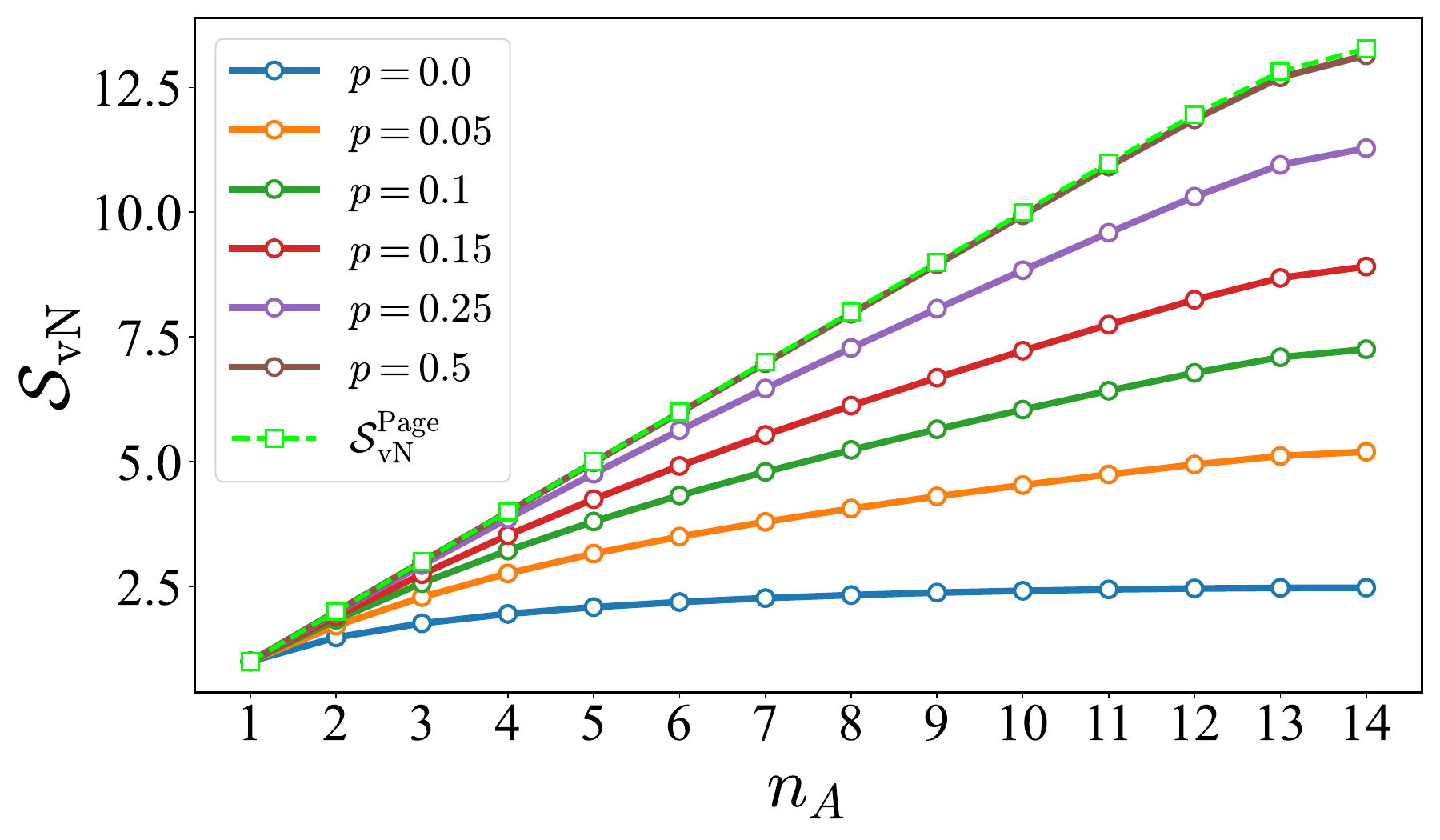}
  \caption{\label{fig:partition_size} Entanglement entropy dependence on subsystem size for sign structure system with $n=28$ qubits and different fractions of minus signs $p$. In all the cases we consider half-filled wave functions with $k = n/2$. The presented dependencies were averaged over 128 instances.}
\end{figure}

Fig.\ref{fig:partition_size} gives the dependence of the entanglement entropy on $n_{A}$ at the fixed system size, $n = 28$. The most interesting feature we observe here is a tiny splitting between entropies obtained for Haar-random state (light green line) and random-sign Dicke states generated with the probability of 0.5 (brown line). Such a splitting takes place only close to the equal bipartition regime.

\begin{figure}[!h]
  \includegraphics[width=\columnwidth]{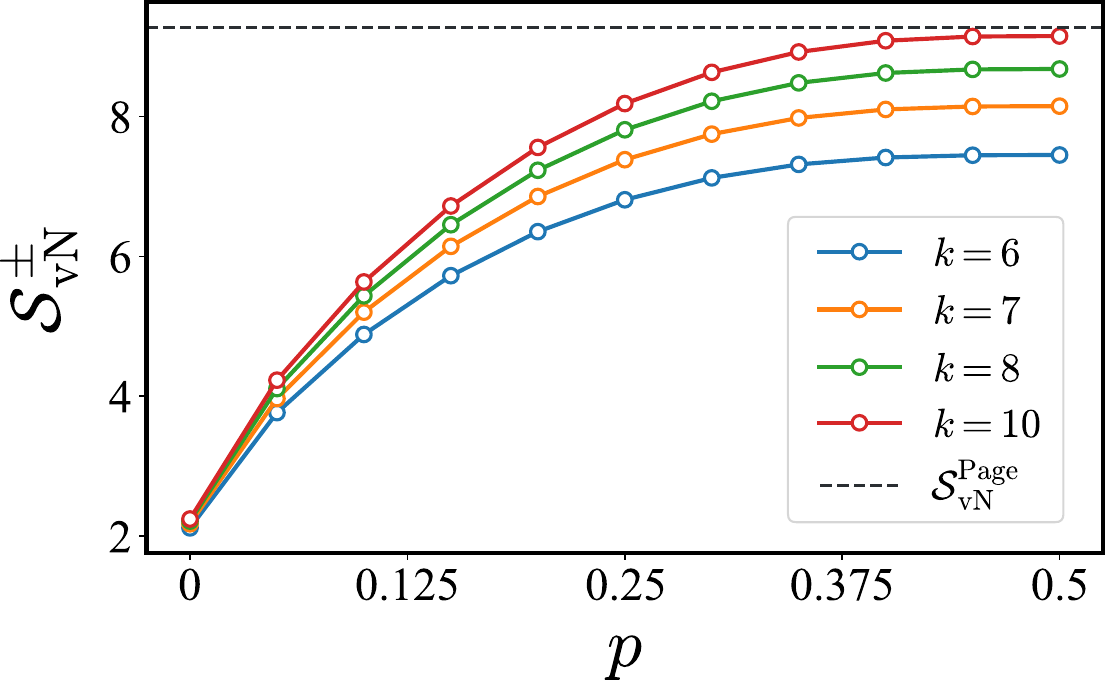}
  \caption{\label{entr} Entanglement entropy of 20-qubit random-sign Dicke states with different $k$ parameters as the function of the probability $p$. The presented dependencies were averaged over 128 instances. The corresponding standard deviation is smaller than the symbol size.}
\end{figure}

The calculations performed for 20-qubit systems, which are presented in Fig.~\ref{entr}\, evidence that the entropy of the Dicke states with $k < n/2$ is sensitive to the negative sign structure in the same fashion as the wave function with half-filling $k=n/2$. 
In all cases, we observe entanglement entropy saturating as the probability approaches 0.5.

\subsection{Invariance of the entanglement to the subsystem choice}
\begin{figure}[!h]
  \includegraphics[width=\columnwidth]{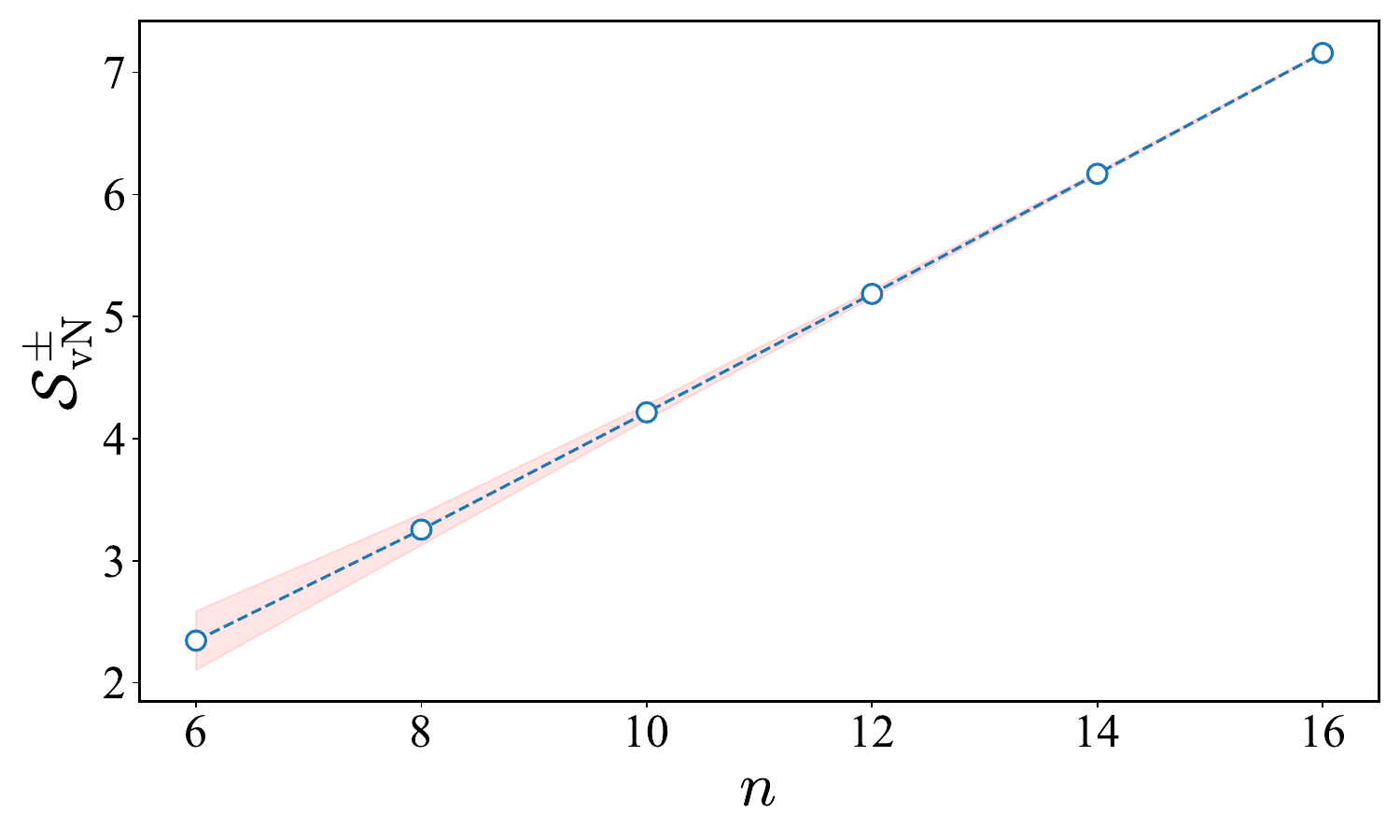}
  \caption{\label{invariant} Calculated entanglement entropies of the random-sign Dicke states, $\mathcal{S}^{\pm}_{\rm vN} (\rho_{A})$ generated at $p= 0.5$ with different number of qubits, $n$. In all the calculations the size of the subsystem A is fixed at $n_{A} = \frac{n}{2}$. For each $n$ we have generated 128 random-sign Dicke states each of which is associated to 10 randomly generated bipartitions. Each blue circle corresponds to the mean entropy averaged over 1280 values and red area denote the standard deviations.}
\end{figure}

The total number of the perfectly balanced bipartitions of the $n$-qubit quantum system is equal to $\frac{n!}{(\frac{n}{2})! (\frac{n}{2})!}$ for even $n$ and coincides with the state space size of a $\Ket{\pm D^{n/2}_{n}}$ wave function. To justify our conclusion on the volume law behaviour of the entanglement of the random-sign Dicke states we have performed the following calculations. First, for each system of 6, 8, 10, 12, 14 or 16 qubits we have generated 128 random-sign Dicke states at $p = 0.5$ by using the procedure described in the section ''Sign structure'' of the main text. Then for each wave function we have randomly generated 10 different bipartitions with $n_{A} = n_{B}$. Thus, for each system size we have calculated 1280 values of $\mathcal{S}^{\pm}_{\rm vN} (\rho_{A})$. Fig.\ref{invariant} gives the mean values of thus computed entropies (blue circles) and its standard deviations (red area). Observing a considerable suppression of the standard deviations as $n$ increases we conclude that $\mathcal{S}^{\pm}_{\rm vN} (\rho_{A})$ is invariant to the choice of the qubits forming subsystem A.

\begin{figure}
  \includegraphics[width=\columnwidth]{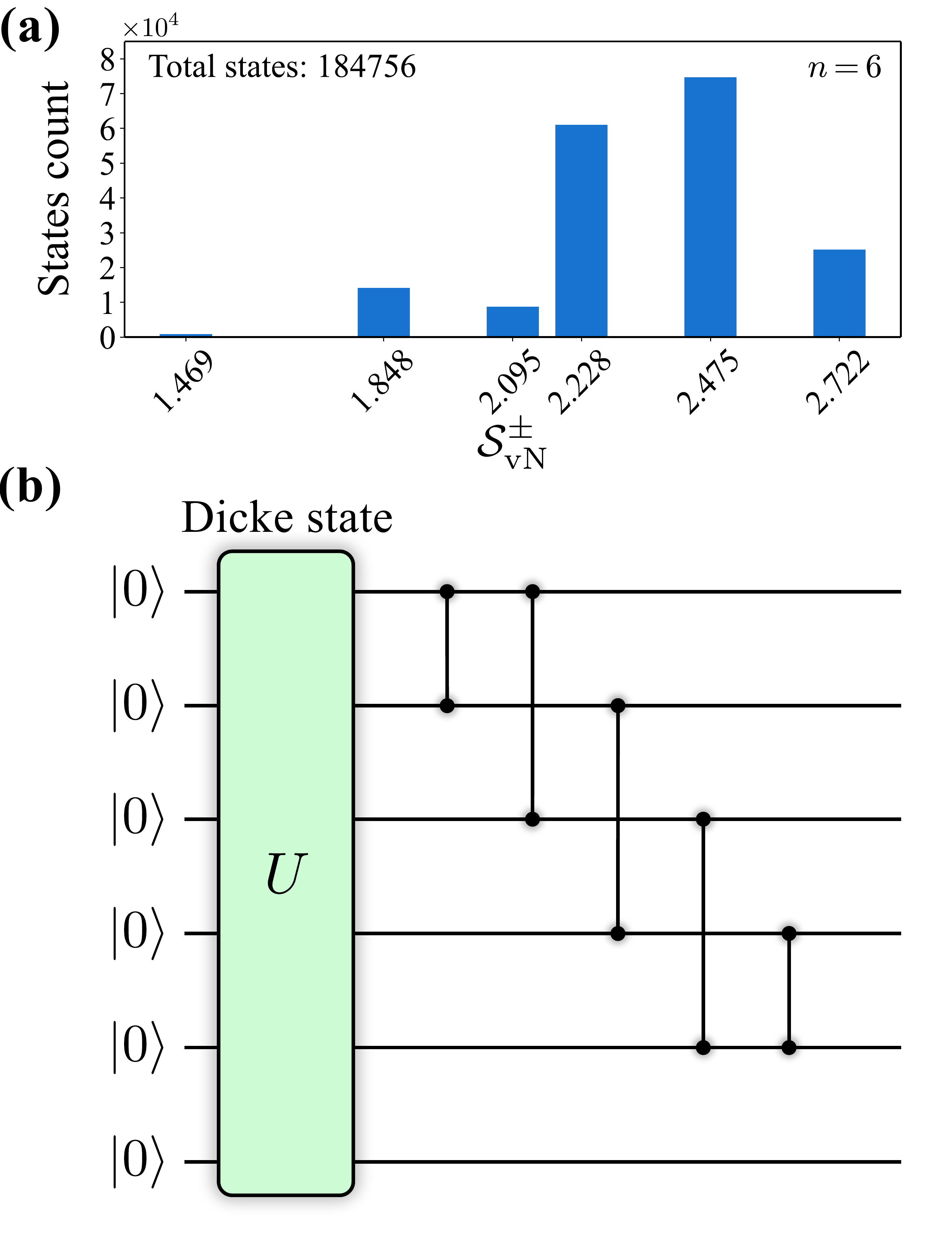}
  \caption{\label{fig:invariant_states} (a) Entropy distribution of all possible signed structure states in 6-qubit system. (b) Example of circuit initializing partition-invariant signed structure state having maximal entropy value.}
\end{figure}

\subsection{A parent Hamiltonian for the 6-qubit highly-entangled Dicke state with sign structure}

In this section we present physical realization of the random-sign Dicke states. As the example we choose the following 6-qubit wave function
\begin{eqnarray}
\Ket{\pm D^{3}_{6}} = \frac{1}{\sqrt{20}} (\Ket{000111} + \Ket{001011} - \Ket{001101} -\Ket{001110} \nonumber \\ 
- \Ket{010011} + \Ket{010101} - \Ket{010110}  
-\Ket{011001} + \Ket{011010}  \nonumber \\  
+ \Ket{011100} -\Ket{100011} - \Ket{100101} + \Ket{100110} + \Ket{101001}  \nonumber \\ - \Ket{101010} 
+ \Ket{101100} + \Ket{110001} + \Ket{110010} - \Ket{110100}  
\nonumber \\ - \Ket{111000}) \nonumber
\end{eqnarray}
that can be created with the quantum circuit presented in Fig.\ref{fig:invariant_states}\,(b). This quantum state is characterized by the maximal entanglement entropy of 2.722 (Fig.\ref{fig:invariant_states}\,(a)) obtained for subsystem that contains the half of the qubits. Importantly, the resulting quantum correlations are invariant to the choice of the subsystem qubits. 

Expressing CZ gates through the Pauli matrices as described in the main text we obtain the following parent Hamiltonian for the ground $\Ket{\pm D^{3}_{6}}$ state 
\begin{eqnarray}
H_{\pm} = -(\sigma^{x}_{0} \sigma^{x}_{1} + \sigma^{y}_{0} \sigma^{y}_{1} + \sigma^{x}_{4} \sigma^{x}_{5} + \sigma^{y}_{4} \sigma^{y}_{5}) \sigma^{z}_{2} \sigma^{z}_{3} \nonumber \\
- (\sigma^{x}_{0} \sigma^{x}_{2} + \sigma^{y}_{0} \sigma^{y}_{2} + \sigma^{x}_{3} \sigma^{x}_{5} + \sigma^{y}_{3} \sigma^{y}_{5}) \sigma^{z}_{1} \sigma^{z}_{4} \nonumber \\
- (\sigma^{x}_{0} \sigma^{x}_{3} + \sigma^{y}_{0} \sigma^{y}_{3} + \sigma^{x}_{2} \sigma^{x}_{4} + \sigma^{y}_{2} \sigma^{y}_{4}) \sigma^{z}_{1} \sigma^{z}_{5} \nonumber \\
- (\sigma^{x}_{0} \sigma^{x}_{4} + \sigma^{y}_{0} \sigma^{y}_{4} + \sigma^{x}_{1} \sigma^{x}_{3} + \sigma^{y}_{1} \sigma^{y}_{3}) \sigma^{z}_{2} \sigma^{z}_{5} \nonumber \\
- (\sigma^{x}_{0} \sigma^{x}_{5} + \sigma^{y}_{0} \sigma^{y}_{5} + \sigma^{x}_{1} \sigma^{x}_{2} + \sigma^{y}_{1} \sigma^{y}_{2}) \sigma^{z}_{3} \sigma^{z}_{4} \nonumber \\
- (\sigma^{x}_{1} \sigma^{x}_{4} + \sigma^{y}_{1} \sigma^{y}_{4}) \sigma^{z}_{3} \sigma^{z}_{5} - (\sigma^{x}_{1} \sigma^{x}_{5} + \sigma^{y}_{1} \sigma^{y}_{5}) \sigma^{z}_{2} \sigma^{z}_{4} \nonumber \\
- (\sigma^{x}_{2} \sigma^{x}_{3} + \sigma^{y}_{2} \sigma^{y}_{3}) \sigma^{z}_{4} \sigma^{z}_{5} - (\sigma^{x}_{2} \sigma^{x}_{5} + \sigma^{y}_{2} \sigma^{y}_{5}) \sigma^{z}_{1} \sigma^{z}_{3} \nonumber \\
- (\sigma^{x}_{3} \sigma^{x}_{4} + \sigma^{y}_{3} \sigma^{y}_{4}) \sigma^{z}_{1} \sigma^{z}_{2}. 
\end{eqnarray}

\bibliography{bibliography.bib}

\end{document}